\documentstyle[12pt]{article}
\input{epsf}
\input psfig.sty
\def\title#1{\relax\vspace*{2cm}{\large{\bf #1}}\par\vspace*{13.5pt}}
\def\author#1{{#1}\par\vspace*{13.5pt}}
\def\affil#1{{\it #1}\par}
\def\abstract{\vspace*{27pt}ABSTRACT\par\relax}
\def\section#1{\par{#1}\par}
\def\subsection#1{\par\underline{#1}\par}
\def\subsubsection#1{\par\underline{#1.}\ \ }

\newenvironment{references}{\section{REFERENCES}\vspace*{.5cm}%
\parindent=0pt\frenchspacing%
\parskip=1pt plus 1pt minus 1pt%
\interlinepenalty=1000\tolerance=400%
\pretolerance=10000\hyphenpenalty=10000%
\everypar={\hangindent=1.6pc}
}{}
\def\kms{km s$^{-1}$}
\def\Msun{M_\odot}
\def\ha{H$\alpha$}

\begin{document}

\title{NUCLEAR-TO-DISK ROTATION CURVES AND\\
MASS-TO-LUMINOSITY RATIO IN GALAXIES
\footnote{Presented at 32nd COSPAR General Assembly, Session E1.2,
1998 July 15-17, Nagoya: To appear in proc.
"AGN-Normal Galaxy Connection", ed. A. Kinney}} 
\author{Yoshiaki SOFUE}
\affil{Institute of Astronomy, University of Tokyo\\
Mitaka, Tokyo 181, Japan}

\abstract

High-resolution nuclear-to-outer rotation curves for
Sb, SBb, Sc, and SBc galaxies show generally a
steep nuclear rise and  flat rotation from the disk to halo.
The high-velocity central rotation indicates massive core within bulges. 
Since this characteristics is common to most galaxies,
the high-velocity central rotation 
cannot be due to a particular orientation of non-circular motion.
Using these rotation curves,
we derive the distributions of surface-mass density, and 
compare directly with observed surface-luminosity distributions. 
The mass-to-luminosity ratio (M/L) remains constant in the outer bulge
and disk, while it increases toward the halo, indicating the massive halo.
In the central region, the M/L also increases steeply toward the center,
reaching by an order of magnitude greater value than the disk value, which
may indicate a massive core of
radius $\sim 100$ parsecs and mass of $\sim 10^9\Msun$.
The core may be an object linking a bulge and
a massive black hole.

\section{1. INTRODUCTION}

Rotation curves are the principal tool to derive the axisymmetric component of
the mass distribution in spiral galaxies for the first-order approximation. 
Higher-order non-axisymmetric fluctuations
such as spiral arms and bar may be superposed, which could, however,
be estimated only through extensive numerical simulations
of velocity fields and gas distributions for individual galaxies,
but is out of the scope of this paper.

Rotation curves in the disk and outer regions have been obtained
based on optical and HI-line spectroscopy (Rubin et al 1980, 1982;
Bosma 1981; Clemens 1995; Persic et al 1996; 
Sofue 1996, 1997; Honma and Sofue 1997).
These rotation curves have been used to estimate the mass distribution
in the disk and halo (e.g., Kent 1987, 1992).
However, the inner-most rotation curves are not necessarily
well investigated yet in high accuracy because of the lack of HI gas,
as well as because of the contamination of strong bulge light
when photographic plates were used. 

In order to derive inner rotation curves, species like the CO molecules
will be most convenient for its high concentration in the
center, for the available high-resolution spectroscopy,
and  for its negligible extinction even toward the nuclear dusty disk. 
We have been, therefore, using high-resolution CO-line data to obtain
most-completely-sampled rotation  curves for nearby galaxies
(Sofue 1996, 1997, Sofue et al 1997, 1998).
In deriving rotation velocity, we have applied the envelope-tracing 
method, which traces the maximum terminal velocities in position-velocity
diagrams along the major axes (Sofue 1996, 1997).
Recent CCD \ha\-line spectroscopy has also made us
available with accurate rotation curves for the inner
regions, because of the larger-dynamic range,
as well as for easier subtraction of the bulge continuum emission
(Rubin et al 1997; Sofue et al 1998).

Fitting of rotation curves by model potentials
have been widely applied to discuss the mass distribution
(e.g., Kent 1987; Sofue 1996).
However, the decomposition of an observed rotation curve
using model potentials is found to be not unique:
For example, an exponential disk and a Plummer's
potential result in nearly identical rotation curves, while their
mass distributions are quite different from each other.

Hence, in order to investigate the mass distribution and the 
mass-to-luminosity ratio in galaxies
without intervened by any potential models,
it will be more convenient to use the surface-mass
density directly calculated from observed rotation curves.
On the other hand, surface-luminosity distributions have
been observed for many galaxies in high accuracy in optical
and near-infrared ranges.
Both the bulge and disk components are known to be well fitted by
exponential law luminosity profiles (de Jong 1996; Heraudeau 1996).

In this paper, we present high-accuracy rotation curves for Sb, SBb,
Sc and SBc galaxies, and discuss their general characteristics.
We derive radial distributions of dynamical surface-mass density, and
compare with the surface photometry.
We also derive the mass-to-luminosity ratio and discuss its radial
variation.

\begin{figure}
\vskip -1cm
\psfig{figure=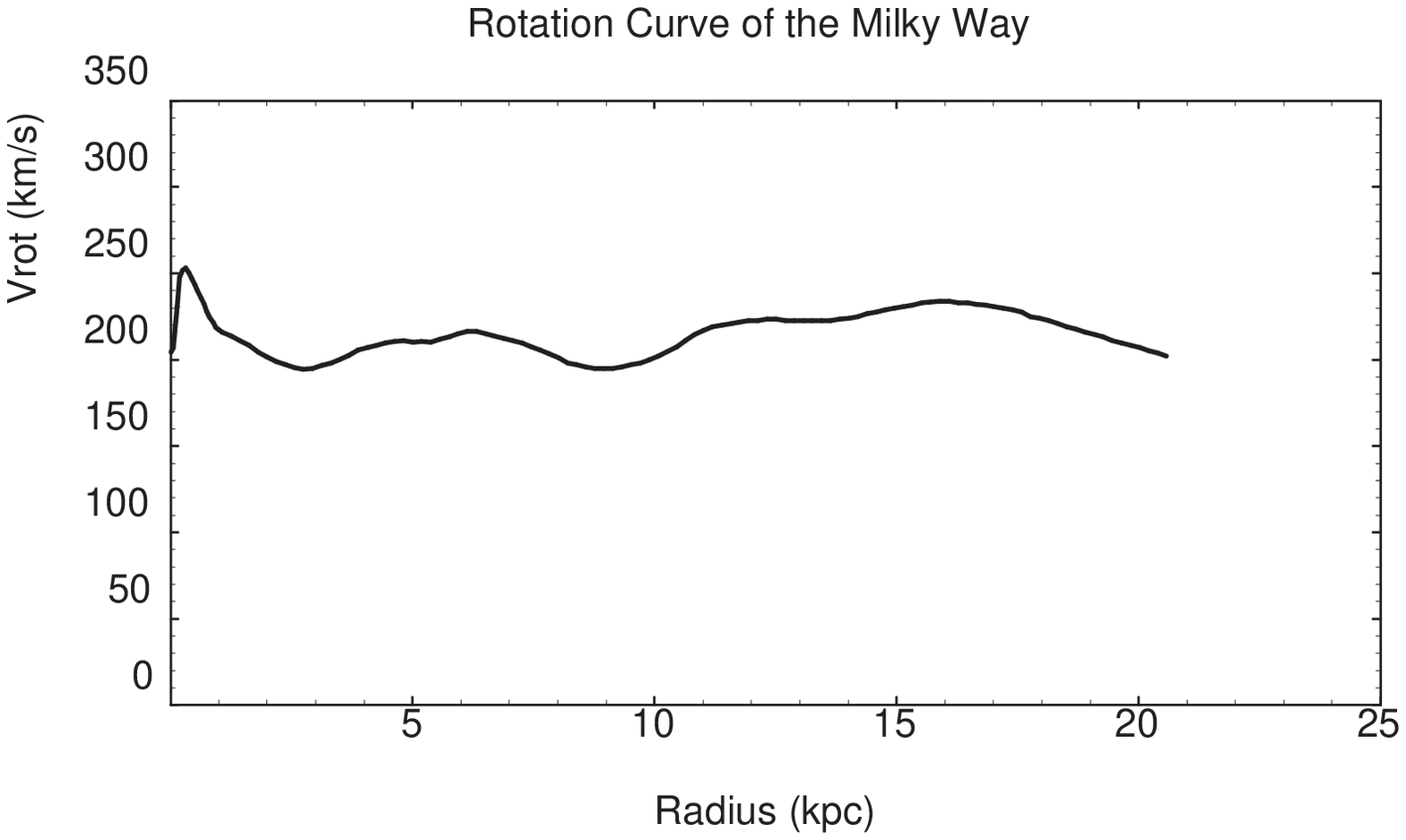,width=18cm,height=9cm}
\vskip 5mm 
\psfig{figure=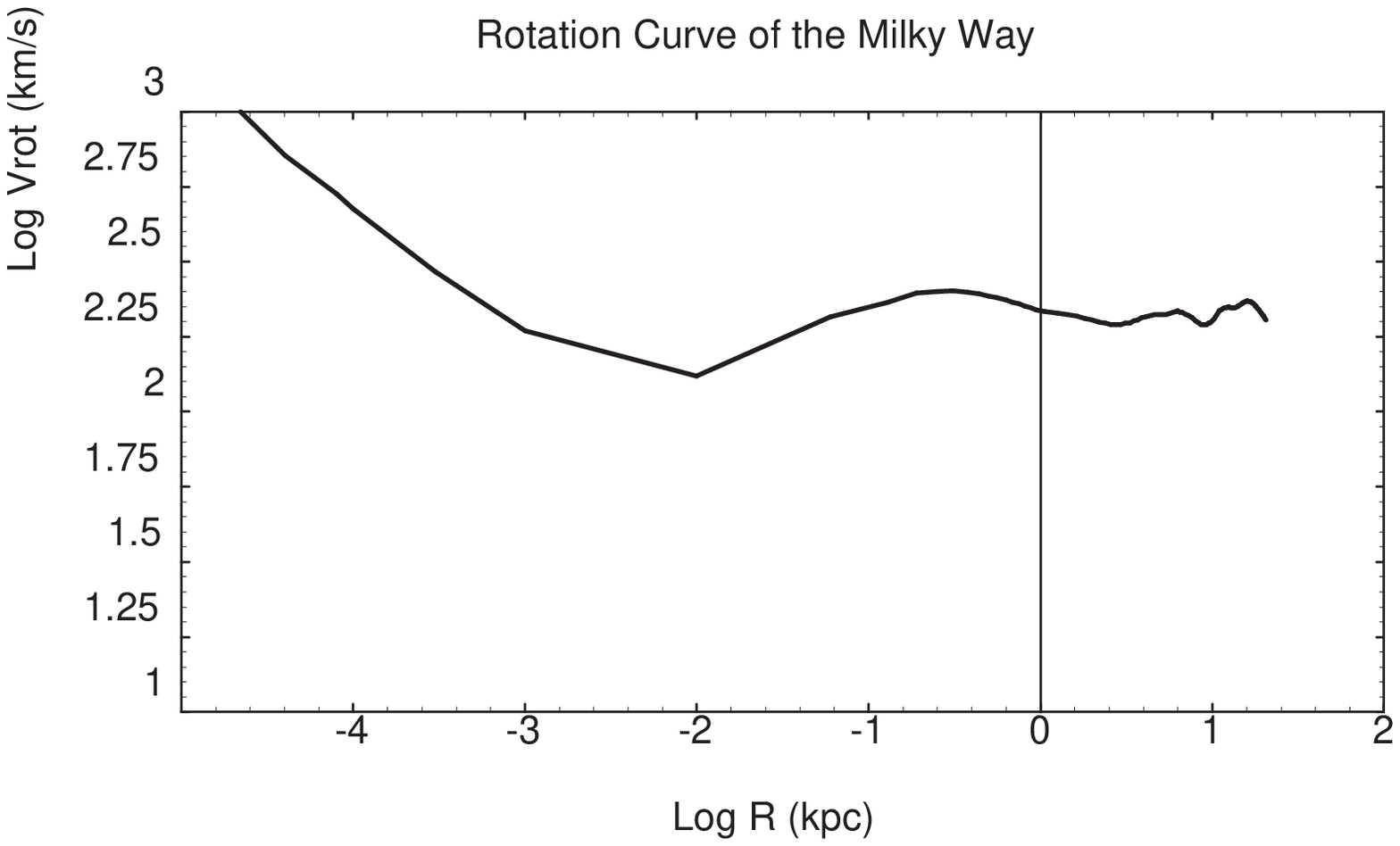,width=18cm,height=9cm}
Fig. 1. (a) "Usual" rotation curve of the Milky Way plotted in
linear scale. (b) "Probably true" rotation curve plotted
in a logarithmic scale, which obeys the Keplerian
near the center due to the massive black hole.
Many galaxies would have similar behavior of rotation
velocity in the center. Note that central zero-velocity
has been adopted merely by a custom to draw a curve
linking positive and negative velocities in both sides of the nucleus 
on PV diagrams.
\vskip 0cm
\end{figure}

\section{2. UNIVERSAL PROPERTIES OF NUCLEAR-TO-OUTER ROTATION CURVES}

\subsection{The Milky Way}

The rotation curve of the Milky Way Galaxy is shown in Fig. 1.
Fig. 1a is a usual plot in a linear scale. 
Genzel et al (1997, 1998) have shown that the velocity dispersion
of stars within the central 10 pc does not decline to zero, but
behaves in a Keplerian fashion, indicating
the existence of a massive black hole.
Combining the velocity dispersion and high-velocities
observed in the CO line emissions, we may raw a "rotation
curve" of the Milky Way as shown in Fig. 1b in logarithmic scales.
Here the velocity dispersion and rotation velocity have been taken
to be identical, both representing the mass distribution.
Since evidences for nuclear massive black holes have been 
accumulated for many other galaxies (Miyoshi et al 1995;
Faber 1998), we may also speculate that the central rotation
curves of galaxies would be more or less similar to that
as shown in Fig. 1b.

\subsection{How to derive rotation curves}

\begin{figure}
\hskip 2.5cm
\psfig{figure=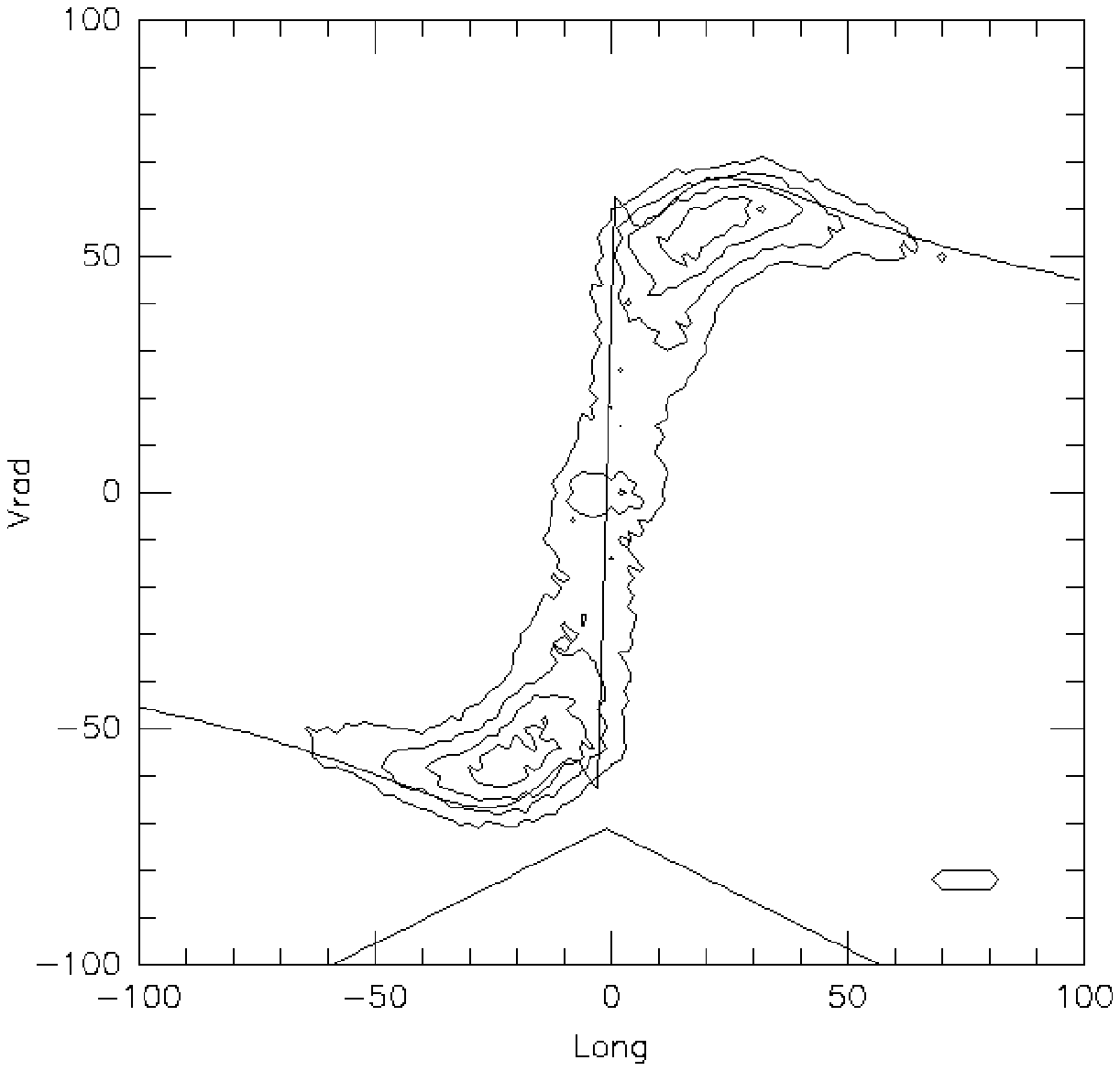,width=12cm,height=9cm}
\vskip 0.5cm
Fig. 2. Simulation of a position-velocity diagram (contours)
from an assumed rotation curve (full line) and
gas distribution (semi-logarithmic density profile at bottom)
for  a finite beam and velocity resolution with some artificial 
noise added.
Coordinates are arbitrary.
The bulge and inner-disk kinematics may be easily missed, if we trace
the peak-intensity positions on PV diagrams.
\end{figure}

Besides the Milky Way, it has been widely believed 
that inner rotation curves behave in a rigid-body fashion
(Rubin et al 1980, 1982; Persic et al 1996).
For example, the bulge and inner bulge components are not
detected in the so-called "universal rotation curves" presented by
Persic et al (1996) based on about a thousand optical and HI rotation
curves observed by Mathewson et al (1996), whose major purpose
was, however, to discuss the dark halo.
These RCs appear to behave very differently from that of the
Milky Way's rotation in the inner disk and bulge regions.

A question may arise, either if the Milky Way is an exception, or
something is missing in the inner parts of 
these "universal" rotation curves.
In order to clarify this, we have simulated position velocity diagrams 
based on an assumed rotation curve and gas distribution
for finite velocity and angular resolutions with some noise.
Fig. 2 shows an example, where the gas distribution is exponential 
and the rotation curve comprises the bulge and disk components.
No bulge component shows up in this simulated PV diagram.
This suggests that the central kinematics may be easily missed, 
if we trace the peak-intensity positions in PV diagrams,
and that a high-resolution PV diagram of species, which
is highly concentrated to the center, is necessary to derive
nuclear rotation curves.

\subsection{Most-completely-sampled rotation curves}

We have been undertaking compilation of high-resolution CO-line
PV diagrams, both from our own observations and from the
literature, and combined with the existing HI and optical
rotation curves.
We have also obtained CCD spectroscopy in the \ha\
and [NII] line emissions of the central regions of galaxies.
We also adopted the 'envelop-tracing method' to derive rotation
curves from the PV diagrams (Sofue 1996, 1997).
In Fig. 3 we show the thus obtained
most-completely sampled rotation curves for Sb and Sc galaxies
(Sofue 1996, 1997; Sofue et al 1997, 1998). 
Dashed curves are for barred galaxies.
In Fig. 4 we show rotation curves for Sb and Sc galaxies separately.
In so far as our sample galaxies, whose central rotation curves 
have been derived from high-resolution CO line data, are concerned,
the rotation characteristics of spiral galaxies is essentially
the same as that of the Milky Way.

\begin{figure}
\hskip 0.0cm
\psfig{figure=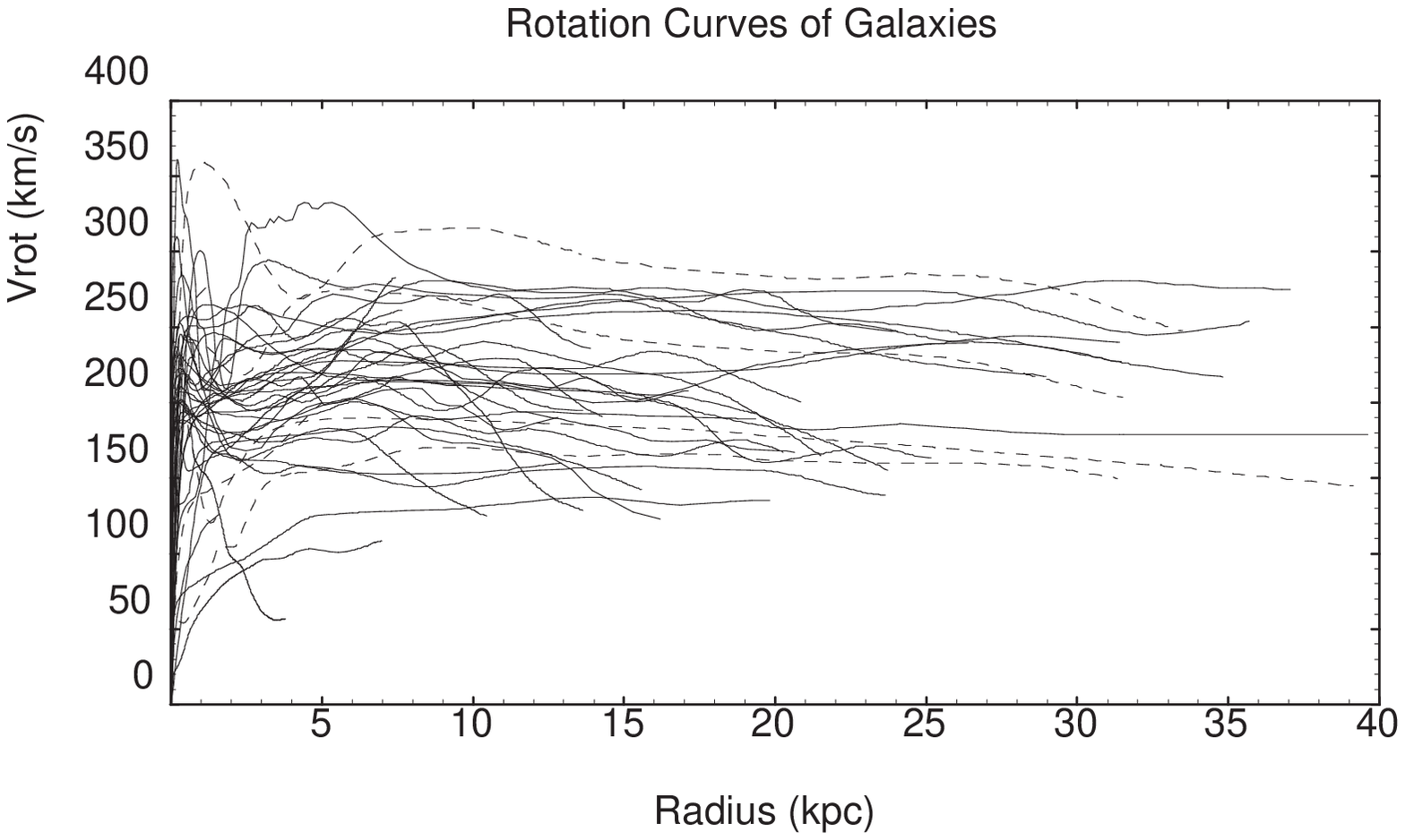,width=18cm,height=9cm}
Fig. 3. Most-completely-sampled rotation curves  of
Sb and Sc galaxies obtained by using CO, \ha\ and HI-line data.
Dashed lines are for barred galaxies.
\end{figure}

\subsection{Sb galaxies}

\begin{figure}
\vskip -3cm
\psfig{figure=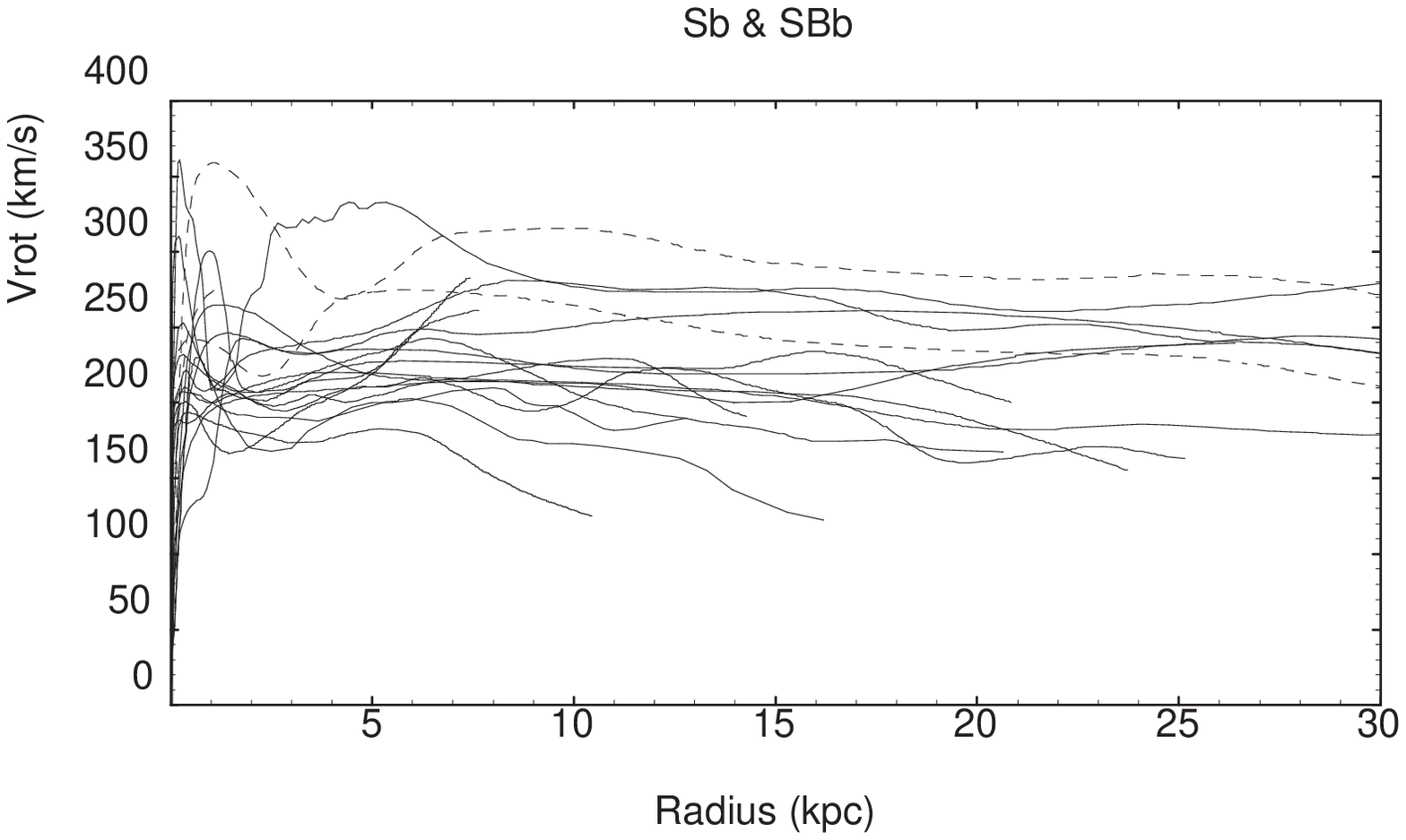,width=18cm,height=9cm}
\vskip 1mm
\psfig{figure=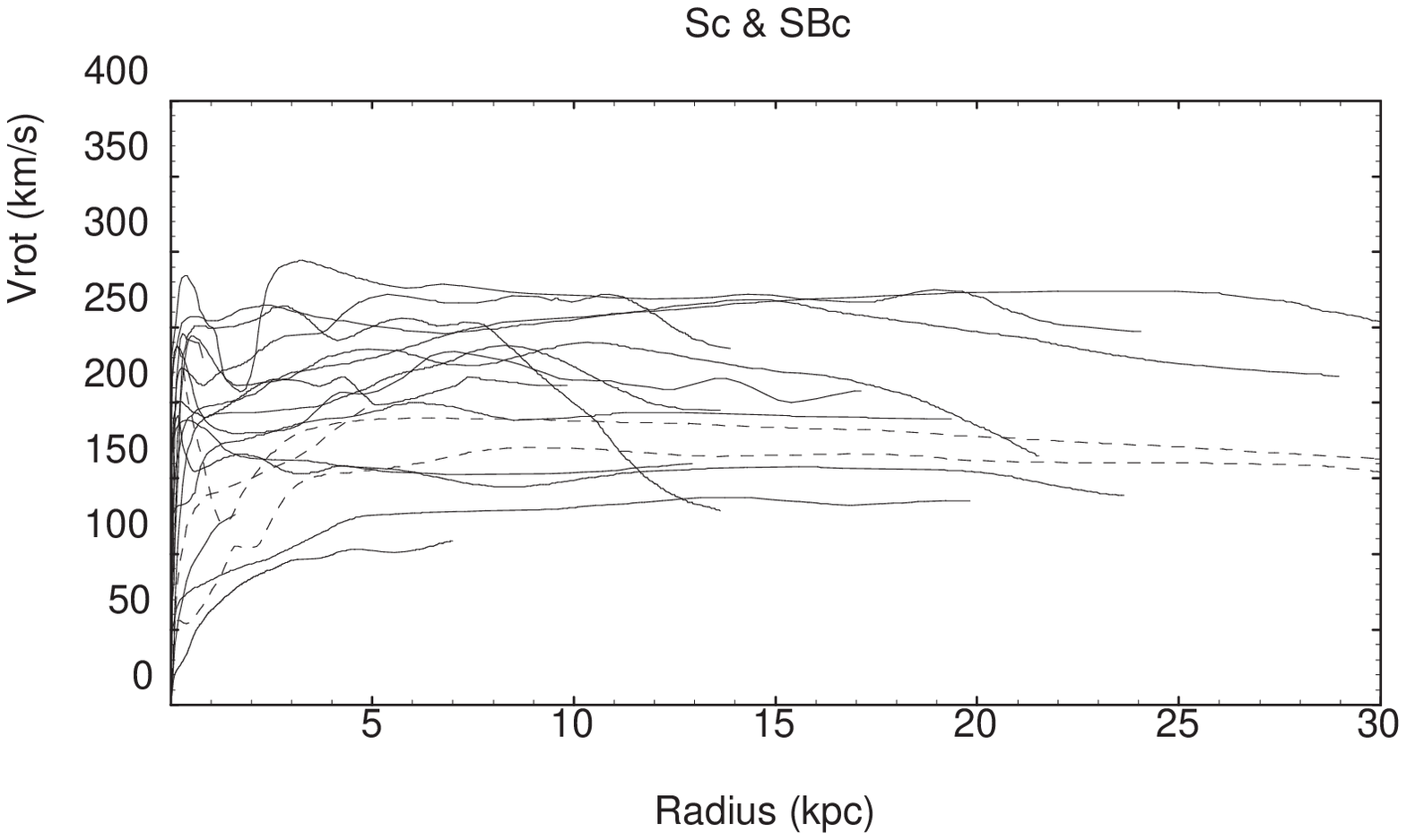,width=18cm,height=9cm}
Fig. 4. (a) 16 Sb (full lines) and 3 SBb (dashed) galaxies.
(b) 16 Sc (full lines) and 3 SBc (dashed) galaxies;
Note that small-mass galaxies, whose maximum disk velocities
are 100 to 200 \kms, tend to show mild rise.
\end{figure}

All Sb galaxies have rotation curves with a very steep
rise in the central 100-200 pc region, often associated 
with a sharp peak at radii $r \sim 100-300$ pc.
The rotation velocity, then, declines to a minimum at $r\sim 1 $ kpc,
and is followed by a gradual rise to a broad maximum at
$r \sim 2-10$ kpc, corresponding to the disk.
The outermost part are usually flat,
while some galaxies like the Milky Way show a declining outer 
rotation (Honma and Sofue 1996, 199).
In general, rotation curves for Sb galaxies are similar to 
that of our Milky Way Galaxy.
There appears no characteristic difference between
normal Sb and barred SBb galaxies. 

\subsection{Sc galaxies}

Sc and SBc galaxies show similar rotation curves,
while having slower velocities than Sb, and the rotation
velocities are more spread from $\sim 100$ to $\sim 200$ \kms\ 
among galaxies.
Massive Sc galaxies show a steep nuclear rise, while
less-massive galaxies have a more gradual rise.
They also have a flat rotation until their outer edges.

\subsection{Barred galaxies}

Some galaxies in Fig. 3 are barred galaxies. 
Typical SBb galaxies are NGC 1097, NGC 1365, and NGC 6674;
SBc galaxies are NGC 3198, NGC 5236, and UGC 2855.
Except that the SBb galaxies have higher rotation velocity than SBc,
there appears no particular difference in their general properties:
The rotation property of the barred galaxies is almost the
same as those for normal Sb and Sc galaxies. 

\subsection{Irregular galaxies}

Rotation curves for irregular galaxies have been obtained
for NGC 3034, NGC 4631 and NGC 4945.
NGC 3034 (M82) shows a very exceptional behavior of rotation:
it has a steep nuclear rise as usual, but decreases smoothly after its
nuclear peak, obeying the Keplerian law.
NGC 4631 is an interacting dwarf amorphous galaxy, showing
a rigid-body increase of rotation velocity.
However, since this galaxy is edge on, it is not clear, if the CO
gas is missing near the nucleus, which might have resulted in a
pseudo rigid-body rotation.
NGC 4945 shows a quite normal rotation with a steep nuclear rise and flat
disk rotation.

\subsection{Universal property: Steep nuclear rise 
and non-zero velocity at center}

Our data in Fig. 3 and 4 are based mainly on CO line data,
which traces the circular rotation of the innermost region of 
galactic disks. 
Galaxies observed at high resolutions with the Nobeyama mili-meter
Array show a very steep nuclear rise.
Moreover, nearer galaxies with higher linear resolution
tend to have a steeper nuclear rise, which would suggest that
farther galaxies might not be resolved of the central
steeper rise.
From these facts we may conclude that
the steep nuclear rise in the central region
is a universal property for Sb and Sc galaxies,
regardless the existence of a bar.
Recent optical CCD observations have also shown
the nuclear rise (Rubin et al 1997; Sofue et al 1998),
in agreement with the CO results.

It is also interesting that the rotation velocity in many galaxies
does not decline to zero at the nucleus.
This indicates that the mass density increases toward the
nucleus more rapidly than expected from exponential distribution 
of surface-mass density.
We mention that the widely adopted zero-velocity at the center
might be merely due to a custom to draw a curve by linking 
positive and negative velocities from both sides of the nucleus. 

The extremely high frequency of galaxies showing
the nuclear rise indicates that the 
high velocity is not due to an end-on view of
non-circular motion by chance.
Note that the probability to look at a bar end-on
is much smaller than that of side-on view, which should result in
a larger probability for apparently slower rotation
than circular velocity.
Therefore,  the mass estimated from the circular assumption
would be even underestimated in many galaxies, if they contain a bar. 

\subsection{Activity and rotation curves}

No particular correlation has been found of the rotation curves
with central activities.  
Our sample includes starbursts (e.g., NGC 253, NGC 1808), 
Seyferts (NGC 1068, NGC 1097),  
LINERs (NGC 3521, NGC 4569, NGC 7331),
galaxies with jets (NGC 3079), and
galaxies with massive black holes (Milky Way, M31, NGC 4258).
Only an exception is the starburst galaxy NGC 3034 (M82),
which shows a 'usual' nuclear rise but is followed by a Keplerian
disk.

\begin{figure}
\vskip -4.0cm
\hskip 2.0cm
\psfig{figure=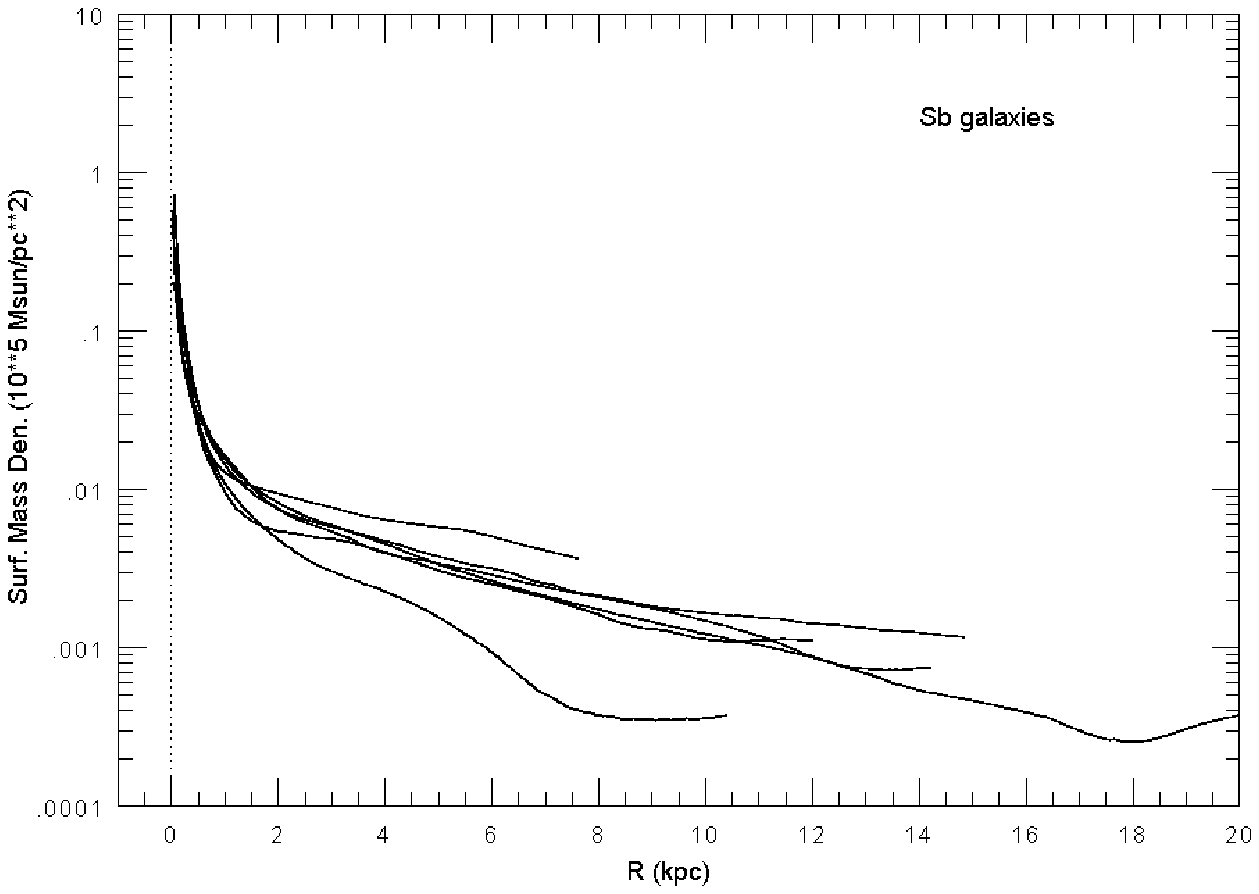,width=12cm,height=9cm}
\vskip 1cm
\hskip 2.0cm
\psfig{figure=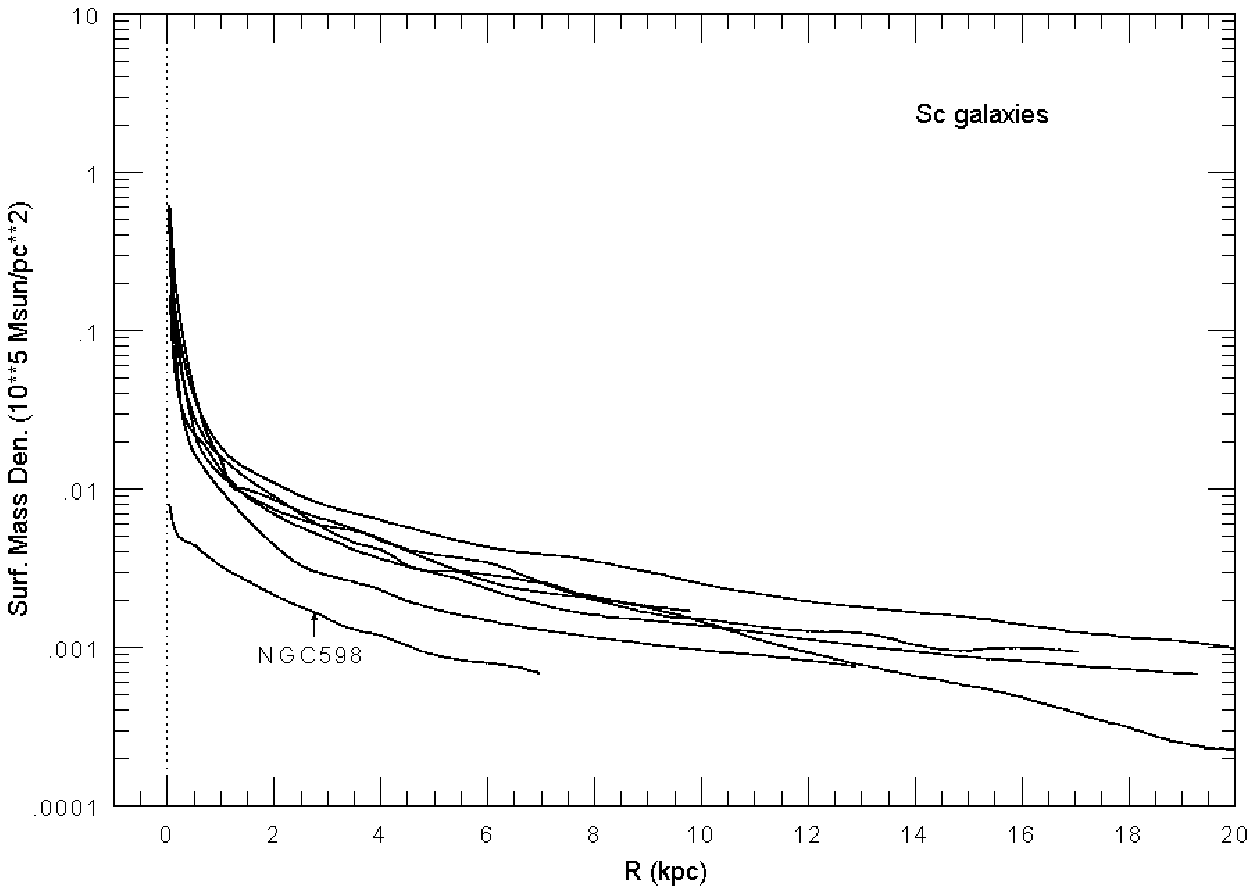,width=12cm,height=9cm}
\vskip 1cm
Fig. 5. Surface mass density of Sb and Sc galaxies as calculated for
disk assumption (Takamiya and Sofue 1998).
\vskip 0cm
\end{figure}

\section{3. MASS DISTRIBUTION AND MASS-TO-LIGHT RATIO}

\subsection{Four mass components}

Using the rotation curves, we have calculated radial 
distributions of surface-mass density (SM), assuming both
spherical symmetry and a flat disk (Takamiya and Sofue 1998).
We found that both assumptions
resulted in a similar mass distributions, differing at most by
a factor of 1.5. 
We stress that a complete set of nuclear-to-outer rotation curve 
is necessary to calculate the SM distribution.
Fig. 5a and b show the thus obtained radial distributions
of surface-mass density for Sb and Sc galaxies for a disk assumption.
The error in the mass estimate is approximately $\pm 20$\%,
which arises from the error in the
rotation velocity,
and the systematic error depending on the disk/sphere assumption
is about a factor of 1.5 at most, as demonstrated in Fig. 6.

The thus calculated surface-mass density is well
represented by four major components:
(1) a nuclear dense core, (2) a bulge,
(3) an exponential disk, and (4) a halo.  
The surface mass density in the inner 1 kpc increases toward
the center according to the bulge.
It shows a particularly steep increase inside 0.3 kpc,
which cannot be represented by an exponential law as indicated from
surface photometry.

\subsection{Common mass distribution regardless the activity}

Sb and Sc galaxies, including  some barred galaxies, 
are found to show similar mass distributions.
Also, the central activities appear to be not directly correlated
with the mass distribution. 
Note that the sample includes galaxies
showing starburst (NGC 3034),
Seyfert (NGC 1068, NGC 2841, NGC 4569), 
LINER (NGC 3521, NGC 4569, NGC 7331),
jets (NGC 3079), or black holes (Milky Way, NGC 4258).

\begin{figure}
\vskip -1.0cm
\hskip 1.0cm
\psfig{figure=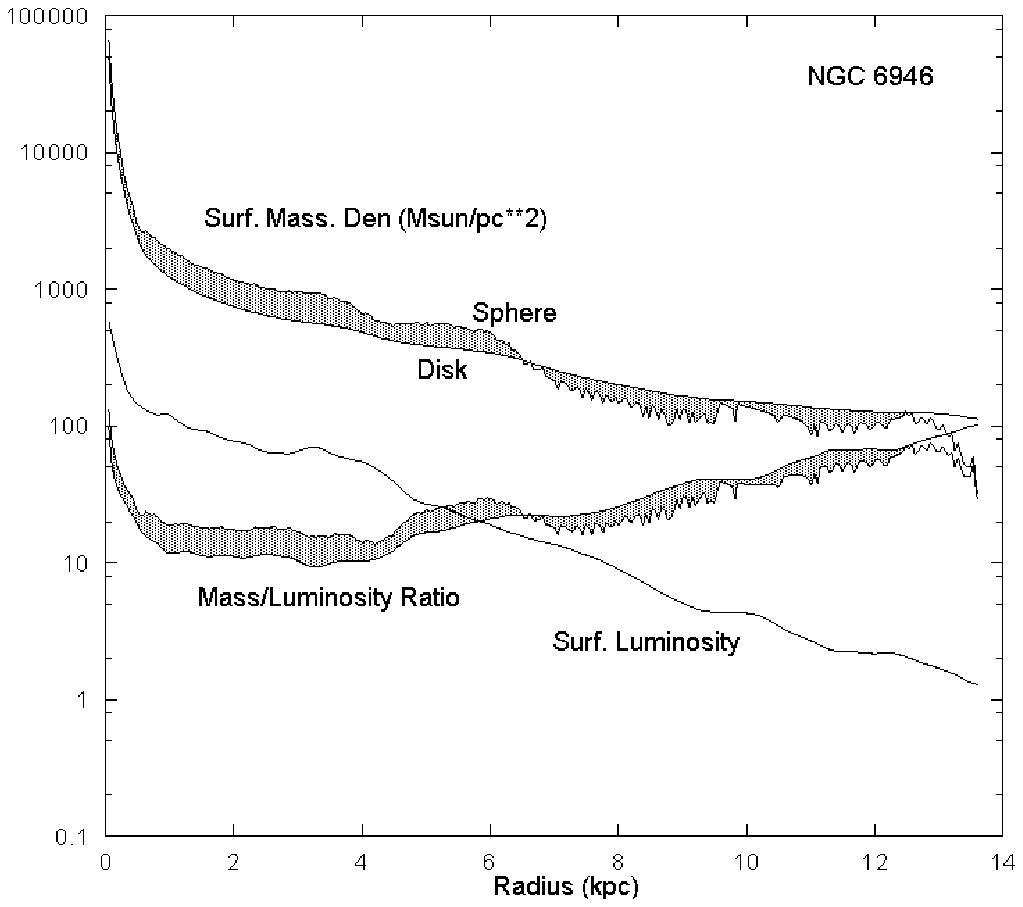,width=15cm,height=9cm}
\vskip 1.0cm
Fig. 6. Radial distributions of surface mass density for a spiral
galaxy NGC 6946 calculated both for disk and spherical assumptions,
indicated by two lines filled with shadows.
The true mass distribution will lie between these two lines.
Observed B-band surface luminosity is shown by the thin line.
The mass-to-luminosity ratio is also shown by two lines
filled with shadow.
The ML ratio in the central region exceeds that of the disk and
outer bulge, suggesting a dark core mass.
It also increases toward the halo due to the dark halo.
\end{figure}

\subsection{Radial variation of Mass-to-Luminosity ratio}

We have calculated the M/L ratio (Takamiya and Sofue 1998)
using the calculated surface-mass distributions and 
observed surface-luminosity data
from the literature (e.g., Kodaira et al 1990; de Jong 1996;
Heraudeau et al 1997).
Fig. 6 plots the obtained distribution of M/L ratio 
for the spiral galaxy NGC 6946. 
The M/L ratio in the disk remains almost constant at $R\sim 0.5$ to 7 kpc.
The outer-disk M/L
increases gradually outward, indicating the massive dark halo.
The normal bulge component, clearly visible in the surface-mass
distribution at $R=0.5-1.5$ kpc,  has almost the same
M/L as the disk, implying that the bulge stars have a similar
M/L to disk stars.
It should be stressed, However, that the M/L increases toward 
the center very steeply.
This may indicate either a dark mass concentration in the central
100 pc, or a significant extinction of the luminosity.
Since NGC 6946 is nearly face on, the latter case would be not
so realistic, suggesting preferably a dark massive core.

\section{4. DISCUSSION}

Many spiral galaxies, for which sufficiently high-resolution
nuclear rotation curves are available, have similar 
rotation curves to that of our Galaxy: 
(1) Steep nuclear rise, or more likely starting from high velocity
(e.g., $c$); 
(2) bulge component; 
(3) broad maximum by the disk; and  
(4) flat halo component.
These correspond to the four mass components: a nuclear core,
a bulge, disk, and massive halo.
The calculated mass distributions for the sample galaxies have
shown no particular correlation with their central activities.

The steep increase of M/L at $R<100$ pc observed for some galaxies
may imply that a galactic bulge contains a 'dark massive core'.
The dark core has a scale radius of $\sim 100$ pc or less, and a mass
$\sim 10^9\Msun$, where the mass-to-luminosity ratio exceeds that of
the mean disk and bulge values by an order of magnitude. 
The origin of the dark core, however, remains open to discussion.
A dark core could be an object linking the galactic bulge and
a massive black hole, having a crucial
implication for the formation and evolution of galactic bulges.

\end{document}